%% file: apssamp.tex
\documentclass[%
 reprint,
 amsmath,amssymb,
 aps,
pra,
]{revtex4-2}

\usepackage{float}
\usepackage{graphicx}
\usepackage{dcolumn}
\usepackage{bm}

\usepackage{color,soul}

\begin{document}

\preprint{APS/123-QED}

\title{Compact Accelerator-Based Production of Carrier-free $^{177}$Lu \\
From 18 MeV $D^+$ on [$^{176}$Yb]Yb$_2$O$_3$}

\author{Austin A. Morris}
 \affiliation{State Key Laboratory of Nuclear Physics and Technology, Peking University}
 \author{Tianhao Wei}
 \affiliation{State Key Laboratory of Nuclear Physics and Technology, Peking University}
 \author{Zhi Wang}
 \affiliation{State Key Laboratory of Nuclear Physics and Technology, Peking University}
 \author{Ying Xia}
 \affiliation{State Key Laboratory of Nuclear Physics and Technology, Peking University}
 \author{Meiyun Han}
 \affiliation{State Key Laboratory of Nuclear Physics and Technology, Peking University}
 \author{Yuanrong Lu}
 \affiliation{State Key Laboratory of Nuclear Physics and Technology, Peking University}
\email{yrlu@pku.edu.cn}

\date{\today}

\begin{abstract} 
We use experimental and simulated excitation functions to estimate the yield of deuteron activations on a [$^{176}$Yb]Yb$_2$O$_3$ target enriched to 99\%. Subsequent calculations are used to determine the production of radiotherapeutic $^{177}$Lu according to a 10 mA, 18 MeV $D^+$ compact linear accelerator. 
 The design comprises a single radio-frequency quadrupole accelerator (RFQ) and seven drift tube linacs (DTLs) that achieve a beam efficiency of 98.5\% over a length of $12\,\text{m}$. Our results show that a 5-day irradiation can yield more than $1$ mg of $^{177}$Lu, exceeding $4.4$ TBq. After a 2 day processing period, it is estimated that the sample will have a radiopurity greater than 99.8\% (carrier-free). Given recent EMA and FDA approvals of $^{177}$Lu-DOTATATE and $^{177}$Lu-PSMA-617, our results confirm the viability of accelerator-based $^{177}$Lu production and provide a promising clinical alternative to reactor-based methods.

\end{abstract}

\maketitle

\input{Introduction.tex}

\bibliography{apssamp}

\end{document}

%% file: introduction.tex
\section{Introduction}

Several studies have reported on the potential of $^{177}$Lu as an in vivo radioisotope useful for the treatment of metastatic lesions and neuroendocrine cancers \cite{dash,nagai,yang}.  $^{177}$Lu is a desirable radioisotope due to its intermediate half-life ($t_{1/2}$ = 6.65 days) and moderate energy release. It beta decays to a stable ground state of $^{177}$Hf (78.6\% of the time), to a 0.25 MeV
excited state (9.2\%), and to a 0.32 MeV excited state (12.2\%). Compared to other therapeutic radioisotopes (such as $^{90}$Y, $^{131}$I, and $^{186}$Re), the localized beta particles produced by $^{177}$Lu have a smaller mean penetration range in tissue of approximately $0.62 \,\text{mm}$ \cite{patell}.
Since the excited states also produce low-energy gamma rays—113 keV (6.6\%) and 208 keV (11\%)—PET image-guided, \textit{theranostic} (therapy and diagnosis) treatments are possible \cite{javier,alliot,vogel,morgan}. Furthermore, $^{177}$ Lu can be combined with other radionuclides, such as $^{67}$Cu and $^{90}$Y, for tandem use \cite{shao}.

For targeted radiotherapy, lutetium can be added to a variety of molecular carriers, antibodies, and peptides to preferentially accumulate in cell receptors over expressed on tumor surfaces \cite{dash, alliot, typ}.  Examples include prostrate-specific membrane antigen (PSMA) and peptide receptor radionuclide therapy (PRRT).  An amino acid transporter protein, PSMA is expressed 100 to 1000 times more in prostate cancer than in normal tissue \cite{ribes,patell,sartor,mendez,justin}. Successful treatments of patients with metastatic lesions and prostrate cancer have led to recent approval by the EMA and FDA of $^{177}$Lu-DOTATATE and $^{177}$Lu-PSMA-617 for neuroendocrine cancers and PSMA-positive metastatic castration-resistant prostrate cancer (mCRPR) \cite{shao,kong,typ,bhar,ribes,sartor,wolfgang,hoj,lisa}. Early results indicate that $^{177}$Lu-based drugs have limited side effects and a better toxicity profile compared to other radionuclides \cite{mendez,wolfgang,justin}. Consequently, the demand for $^{177}$Lu has increased in recent years, with a particular focus on developing production facilities that are better equipped to meet treatment requirements.

In this study, we present the general design of an 18 MeV $D^+$ linear accelerator (linac) optimized for the production of high purity $^{177}$Lu.  We develop a model based on experimental \cite{nagai,herman,man}  and simulated \cite{koning} excitation functions to optimize the target width and yield of deuteron activations from a Yb$_2$O$_3$ target enriched to 99\% $^{176}$Yb. 
Based on our accelerator and target parameters, the expected radiopurity of carrier-free $^{177}$Lu is estimated to exceed 99.8\% after a 5-day irradiation.  Longer irradiations offer similar radiopurity and produce 1 to 2 mg of $^{177}$Lu. Compared to previous studies using reactor-based production, we show that accelerator-based production generates less target burn-up, reduced waste, higher radiopurity, and competitive overall yield. Since an accelerator facility is generally safer and easier to maintain than a high flux reactor, the design is expected to better meet therapeutic production demand.

\section{Accelerator Design}

Currently, $^{177}$Lu is most commonly produced by neutron activation in a high flux reactor \cite{dash, knapp}. This involves either irradiating an ordinary lutetium target or an enriched $^{176}$Yb target (12.9\% natural abundance), giving the reactions
\begin{eqnarray}
^{176}\text{Lu}\, + \,n\, &\longrightarrow& \, ^{177}\text{Lu}\, + \,{\gamma} \;\;\text{(direct),} \\ 
^{176}\text{Yb}\, + \,n\, &\longrightarrow& \, ^{177}\text{Yb}\, + \,\gamma \;\;\text{(indirect)}.
\end{eqnarray}
In the indirect route, $^{177}$Yb ($t_{1/2}$ = 1.91 h) is produced, which quickly beta-decays into $^{177}$Lu. The drawbacks of reactor-based production include the high thermal neutron flux required ($10^{14}$ n/cm$^2$/s) as well as the 2.6\% natural abundance of $^{176}$Lu ($t_{1/2} \sim 4 \times 10^{10}$ y). For both direct and indirect production routes, high neutron activation cross sections lead to significant
target burn-up and radioactive waste, which also reduces the specific activity of $^{177}$Lu \cite{bhar,kp}. The waste products include $^{174\text{g}}$Lu ($t_{1/2} = 3.31$ y), $^{174\text{m}}$Lu ($t_{1/2} = 120.9$ d),
$^{176\text{m}}$Lu ($t_{1/2} = 3.66$ d), and $^{177\text{m}}$Lu ($t_{1/2} = 160.4$ d) \cite{tendl}.  Due to these contaminants, target waste and patient urine must be disposed of in a dedicated waste management
facility \cite{javier}.  Large amounts of stable $^{175}$Lu in the target also reduce the specific activity of the desired $^{177}$Lu product.

By comparison, accelerator-based production methods generate one-tenth as much nuclear waste, require lower maintenance costs (including decommissioning), and have a better safety profile \cite{yiwei}. Proton beams are inadequate for lutetium production as a result of small Yb/Lu activation cross sections. However, deuteron beams, with moderate stopping power and high neutron-stripping cross sections, provide an ideal choice \cite{tark2}.

\begin{figure}[htbp]
    \centering
    \hspace*{-0.4cm}
    \includegraphics[width=9.2cm]{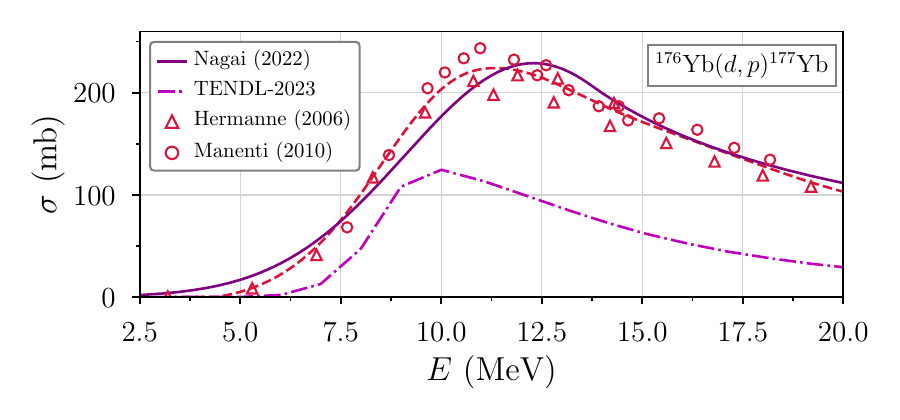}
\vspace*{-0.8cm}
\caption{\label{cs_data} 
Experimental \cite{nagai,herman,man, janis} and simulated \cite{tendl} $^{176}$Yb($d,p$)$^{177}$Yb cross section data. Due to the large underestimation by TENDL-2023 for this reaction, we instead use an interpolation of the Hermanne (2006) and Manenti (2010) data for subsequent calculations.
}
\label{fig:cs}
\end{figure}

Similar to before, $^{177}$Lu is produced indirectly by deuterons through
\begin{eqnarray}
^{176}\text{Yb}\, + \,d\, &\longrightarrow& \, ^{177}\text{Yb}\, + \,p \;\;(Q=3.34 \text{ MeV})
\end{eqnarray}
reactions, at a rate about nine times greater than for direct $^{176}$Yb($d,n$)$^{177}$Lu reactions \cite{kambali}. The $^{176}$Yb($d,p$)$^{177}$Yb cross section begins to increase around $5$ MeV before reaching a peak of approximately $230$ mb at $12.5$ MeV, as shown in Figure \ref{fig:cs}.  

For higher deuteron energies, $^{174g+m}$Lu production cross sections increase \cite{shao}.
Hence, a higher purity of $^{177}$Lu is obtained for deuteron beam energies $E<20$ MeV \cite{nagai}. In this range, metastable $^{176m}$Lu is produced but decays rapidly, while the production of longer-lived $^{177m}$Lu is negligible due to its high spin value $J^\pi=23/2^-$ \cite{tark1}.

\begin{figure*}[htbp]
    \centering
    \hspace*{-0.15cm}
    \includegraphics[width=18.1cm]{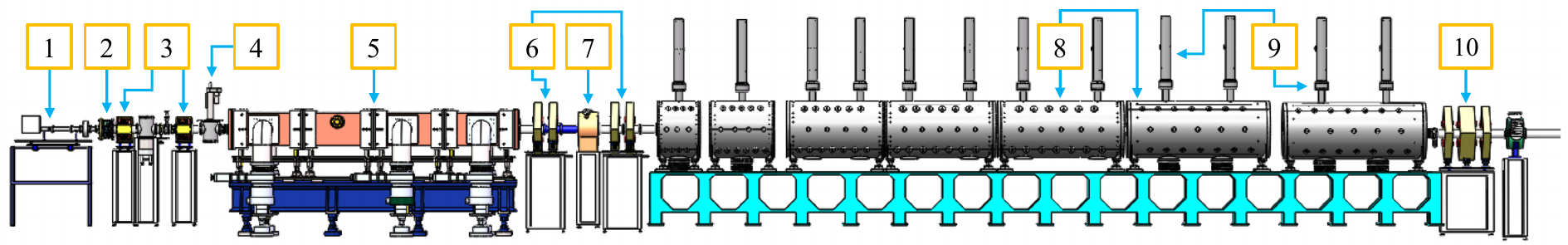}
\caption{\label{RFQ-DTL} 
Side-view schematic of the RFQ-DTL components, including water cooling ports, vacuum pumps, power couplers, and tuners. Labels correspond to: (1) microwave waveguide,
(2) ECR (Electron Cyclotron Resonance) $D^+$ ion source,
(3) solenoids,
(4) emittance measuring device,
(5) RFQ accelerator,
(6) doublet magnets,
(7) buncher,
(8) DTLs,
(9) RF power couplers, and
(10) triplet magnets.
}
\end{figure*}

To produce $^{177}$Lu with high purity and high yield, we designed a 10 mA, 17.9 MeV $D^+$ linear accelerator (linac) \cite{xia}. Our design consists of a 0.02 MeV injector, 3.8 m radio-frequency quadrupole (RFQ) accelerator, and seven drift tube linacs (DTLs). For the RFQ, a 2.1 MeV beam is generated in continuous wave (CW) mode with a simulated transmission efficiency of 98.63\% \cite{wei1}.  The RFQ requires four 1200 L/s vacuum pumps and one 440 L/s ion pump to maintain a vacuum pressure of $10^{-6}$ Pa.
For the high-energy beam section, each DTL consists of a power coupler and six tuners \cite{wei2}.  A crossbar H-mode (CH) cavity increases the ion energy across all seven DTLs from $2.11-17.92$ MeV over 8.17 m. The transverse focusing between cavities is made by quadrupole doublet and triplet magnets depicted in figure \ref{RFQ-DTL}. The beam transmission in each CH-DTL is nearly 100\%, while the entire line efficiency is 98.5\%, due mostly to particles lost at low energy in the RFQ bunching section. The RFQ-DTL system operates in CW mode at a frequency of 162.5 MHz, for the parameters given in tables \ref{tab:tab1} and \ref{structure}.

\begin{table}[htpb]
\begin{ruledtabular}
\scalebox{0.95}{
\begin{tabular}{lc}
  \multicolumn{1}{l}{Accelerator parameters} \\
  \colrule
  \rule{0pt}{2.8ex}
  \!\!Ion & $D^+$\\[2pt]
  Beam energy (MeV) & $17.9$\\[2pt]
  Current (mA) & $10$\\[2pt]
  Frequency (MHz) & $162.5$\\[2pt]
  Duty factor (\%) & $100$\\[2pt]
  Transmission efficiency (\%) & $98.5$

\end{tabular}
}
\end{ruledtabular}
\caption{\label{tab:tab1}%
General parameters of the CW mode RFQ-DTL accelerator design, optimized for $^{177}$Lu production \cite{xia}.}

\vspace{0.2cm}
\begin{ruledtabular}
\scalebox{0.95}{
\begin{tabular}{lcc}
 \multicolumn{1}{l}{Structure} & Length (m) & Energy interval (MeV) \\
  \colrule
  \rule{0pt}{2.8ex}
  \!\!RFQ & $3.8$ & $0.02-2.11$\\[2pt]
  DTL 1 & $0.52$ & $2.11-3.28$ \\[2pt]
  DTL 2 & $0.81$ & $3.28-4.92$ \\[2pt]
  DTL 3 & $1.01$ & $4.92-7.03$ \\[2pt]
  
  DTL 4 & $1.26$ & $7.03-9.37$ \\[2pt]
  DTL 5 & $1.42$ & $9.37-12.01$ \\[2pt]
  DTL 6 & $1.51$ & $12.01-14.86$ \\[2pt]
  DTL 7 & $1.64$ & $14.86-17.92$
\end{tabular}
}
\end{ruledtabular}
\caption{\label{structure}%
Structural parameters of the RFQ-DTL design.  The overall energy interval is 17.9 MeV and the total length of the accelerator is 11.97 m \cite{xia}.}
\end{table}

\begin{figure}[htbp]
    \centering
    \hspace*{-0.4cm}
    \includegraphics[width=9.5cm]{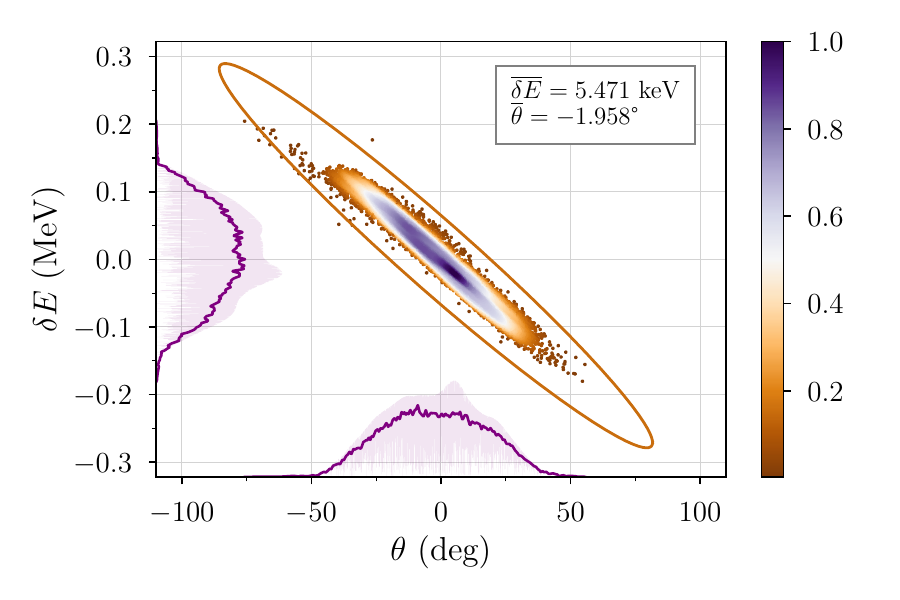}
\vspace*{-0.8cm}
\caption{\label{phase} 
Energy-phase diagram for the 17.9 MeV $D^+$ beam, which is nearly mono-energetic. The results are obtained from output simulated using TraceWin \cite{tracewin}.}
\end{figure}

The beam output was simulated using a TraceWin \cite{tracewin} Monte Carlo simulation of 10,000 particles.  Figure \ref{phase} plots a Gaussian kernel density estimation of particle energy and angle variations, which average 5.47 keV and -1.96°, respectively. The covariance error ellipse is given 4.5 standard deviations from the mean value, corresponding to a $\chi^2$ value $>99.9$\%.  The x-z and y-z filtered flux values indicate a strongly mono-energetic beam profile.

 \section{Target Optimization}

For an ytterbium target,
metallic $^{176}$Yb (6.9 g/cm$^3$) is separated from silicate mineral gadolinite and synthesized into crystalline Yb$_2$O$_3$ (9.2 g/cm$^3$) \cite{Chem}.  Yb$_2$O$_3$ 
is used as a target instead of pure Yb due to its better chemical and thermal stability, as well as its higher melting point (2,355 °C compared to 819 °C) \cite{javier, Chem}. Although natural ytterbium has an abundance of only 12.6\% $^{176}$Yb, it may be enriched to more than 97\% \cite{dash,knapp}. A recent study by Yang \textit{et al.} reported pilot-scale production of $^{177}$Lu from a 99.33\% [$^{176}$Yb]Yb$_2$O$_3$ enriched target purchased from Rosatom \cite{yang}.  Following irradiation, a Yb$_2$O$_3$ target can be dissolved in dilute mineral acid. Since Yb ions exist in a +2 oxidation state, forming an insoluble sulfate, micro amounts of +3 Lu may be separated from macro amounts of Yb, and synthesized into water-soluble LuCl$_3$ \cite{dash,Chem}. Table \ref{tab:tab2} gives the material properties of natural Yb and Yb$_2$O$_3$ used in later calculations.

\begin{table}[htpb]
\begin{ruledtabular}
\scalebox{0.95}{
\begin{tabular}{lcc}
 \multicolumn{1}{l}{Target properties} & {\text{Yb}} & {$\text{Yb}_2\text{O}_3$}\\
  \colrule
  \rule{0pt}{2.8ex}
  \!\!\!\! Molecular weight (g/mol) & 173.04 & 394.08 \\[2pt]
   Density (g/cm$^3$) & 6.9 & 9.2\\[2pt]
   Melting point (°C) & 819 & 2355\\[2pt]
   Specific heat capacity (J/g-K) & 0.155 & n.a.\\[2pt]
   Thermal conductivity (W/m-K) & 38.5 & n.a. \\[2pt]
   Mean excitation potential (eV) & 684 & 512.4\\[2pt] 
   Range of 17.9 MeV D$^+$ (mm) & 0.75 & 0.49
\end{tabular}
}
\end{ruledtabular}
\caption{\label{tab:tab2}%
General properties of natural Yb and molecular Yb$_2$O$_3$ targets \cite{Chem}.  The range of 17.9 MeV deuterons is calculated using equation \ref{range}.}
\end{table}

Unlike neutron activation, for which sharp resonances provide large cross section variations, charged particle activations are straightforward to calculate. The average energy lost per unit path-length, or electronic stopping power of a heavy ion in a dense medium is given by  
\begin{equation}
    -\frac{dE}{dx} = \sum_i N_i \sum_n E_{ni} \,\sigma_{ni},
\label{SP1}
\end{equation}
\noindent where $N_i$ is an atom or molecule of type $i$ that is excited to an energy level $E_{ni}$ above its ground state, for a corresponding inelastic cross section $\sigma_{ni}$ \cite{fano}.  Applying a first-order Born approximation, equation (\ref{SP1}) transforms into the relativistic Bethe-Bloch equation
\begin{eqnarray}
    -\frac{dE}{dx} = \frac{4\pi m_\text{e} c^2 r_\text{e}^2 \, n_\text{e}z^2}{\beta^2}    \left\{\ln\left[\frac{2m_\text{e} c^2 \beta^2}{\langle I \rangle(1-\beta^2)}\right] -\beta^2 \right\}
\label{SP}
\end{eqnarray}
\noindent where the electron density of the medium is $n_\text{e}$, the atomic number of the ejected ion is $z$, and the mean excitation potential is $\langle I \rangle$.  Neglecting the shell and density corrections, which are small for intermediate deuteron energies, and using the fact that
\begin{eqnarray}
    -\ln{\left(1-\beta^2\right)} -\beta^2 = \frac{\beta^4}{2}+\frac{\beta^6}{3} +\dots<<1,
\end{eqnarray}
equation (\ref{SP}) becomes
\begin{eqnarray}
    S = \frac{\omega z^2}{\beta^2}    \ln\left( \kappa \beta^2 \right),
\label{SP_approx}
\end{eqnarray}
for $S=-dE/dx$,  $\omega = 4\pi n_\text{e} m_\text{e}c^2 r_\text{e}^2$, and $\kappa = 2m_\text{e} c^2/ \langle I \rangle$ \cite{morris}.  

Since the mean excitation potential for Yb$_2$O$_3$ molecules is not readily available, we estimate it by using Bragg's formula, 
\begin{eqnarray}
    \ln{\langle I \rangle} \sim \frac{\sum_i \,N_i \, Z_i \ln{I_i}}{\sum_i \,N_i \, Z_i}.
\label{Bragg}
\end{eqnarray}
The mean excitation energy is 684 eV for ytterbium atoms and 95 eV for oxygen atoms \cite{NIST}, giving $\langle I \rangle \sim 512.4$ eV for Yb$_2$O$_3$. From equation 
\ref{SP_approx}, the average range of an ion can be estimated using the continuously slowing-down approximation
\begin{eqnarray}
    R  = \int_0^{E} \!\frac{dE'}{S(E')} =\frac{m_0 c^2}{2 \omega z^2} \!\int_0^{{\beta}^2} \!\!\!\frac{\beta'^2 \,d\beta'^2}{(1-\beta'^2)^{3/2}\ln(\kappa \beta'^2)}.
\label{range}
\end{eqnarray}

\begin{figure}[htbp]
    \centering
    \hspace*{-0.25cm}
    \includegraphics[width=9cm]{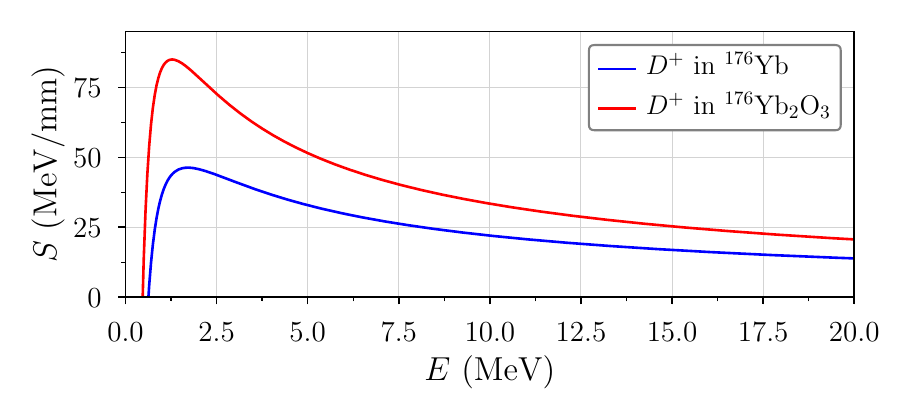}
    
    \vspace*{-0.75cm}
    \hspace*{-0.25cm}
    \includegraphics[width=9cm]{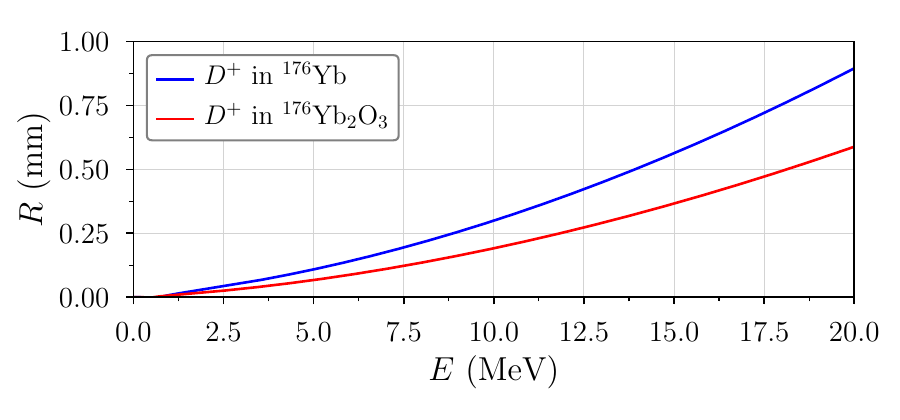}
\vspace*{-0.8cm}
\caption{\label{sp_and_range} 
\textit{top.} Stopping power plots for deuterons in Yb and Yb$_2$O$_3$ targets. \textit{bottom.} Range plots corresponding to ions that slow to rest, in the interval $E\rightarrow 0$ MeV (i.e. $R(E)<T$).
}
\end{figure}

Compared to ranges simulated by SRIM, equation \ref{range} generally exhibits close agreement, with differences of less than 5 $\mu$m \cite{kambali}. Figure \ref{sp_and_range}
plots equations \ref{SP_approx} and \ref{range} for deuterons in pure ytterbium and Yb$_2$O$_3$ targets, consistent with results obtained in references \cite{kambali,shao,javier}. 
Owing to its higher molecular density and lower mean excitation energy, Yb$_2$O$_3$ offers greater stopping power to $D^+$ ions than does pure Yb, resulting in a shorter range of travel. For estimating the yield of a reaction of type $i$ with cross section $\sigma=\sigma^{\{i\}}$, it is convenient to introduce the range-averaged cross section
\begin{eqnarray}
    \left\langle \sigma \right\rangle = \frac{\displaystyle\int_{0}^E \frac{\sigma(E')} {S(E')} \,dE'}{\displaystyle\int_{0}^E \frac{dE'}{S(E')}}  
    \simeq \frac{\displaystyle\int_{0}^E {\sigma(E')} \, E'\,dE'}{\displaystyle\int_{0}^E E'\,{dE'}}.
\label{ran_avg}
\end{eqnarray}
Since the log term in equation \ref{SP_approx} varies slowly with energy $E \sim mc^2\beta^2/2$, it can be treated as a constant in equation \ref{ran_avg}, implying that the range-averaged cross section is nearly independent of the material stopping power. Thus, for thin targets, or those in which ions do not impart all of their energy on average, beam straggling may be neglected \cite{shao}. 

\begin{figure*}[htpb]
    \centering
    \begin{minipage}
    {0.3\textwidth}
        \centering
        \hspace*{-3.705cm}
        \includegraphics[width=9.2cm]{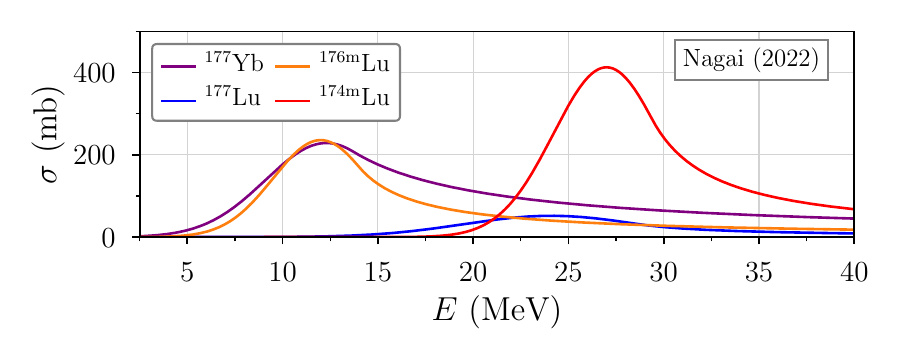}
    \end{minipage}
    \begin{minipage}
    {0.3\textwidth}
        \centering
        \hspace*{-0.315cm}
        \includegraphics[width=9.2cm]{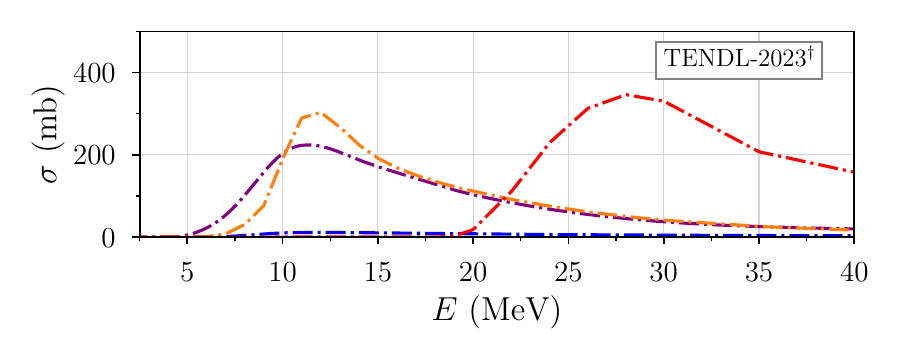}
    \end{minipage}
    
    \centering
        \vspace*{-0.3cm}
    \begin{minipage}
    {0.3\textwidth}
        \centering
        \hspace*{-3.65cm}
        \includegraphics[width=9.2cm]{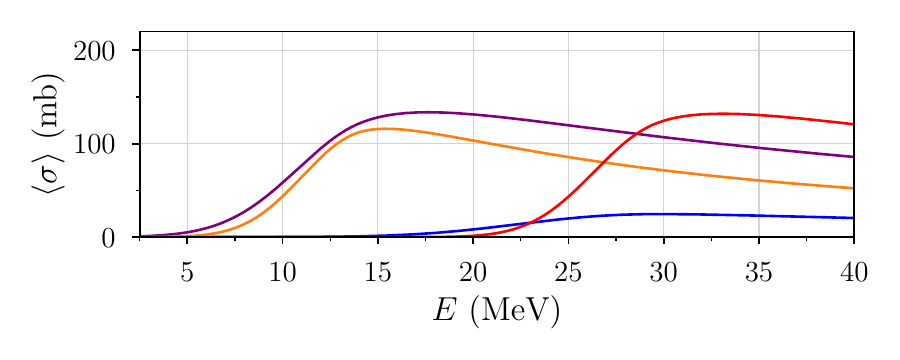}
    \end{minipage}
    \begin{minipage}
    {0.3\textwidth}
        \centering
        \hspace*{-0.26cm}
        \includegraphics[width=9.2cm]{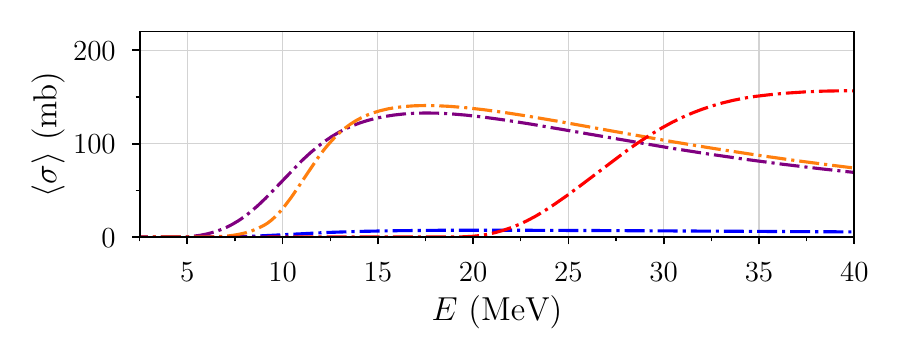}
    \end{minipage}
\vspace*{-0.3cm}   
\caption{\label{fig:cross_sections} 
\textit{top.} Short-lived radioisotope excitation functions given by Nagai \textit{et al.} \cite{nagai} and TENDL-2023$^\dag$ \cite{tendl}, in which we substituted the less accurate TENDL $^{176}$Yb($d,p$)$^{177}$Yb cross section with data interpolated from \cite{man,herman}. \textit{bottom.} Range-averaged cross sections for particles that slow completely to rest (i.e. $R(E)<T$), where $E$ is the initial energy of a monoenergetic beam.}
\end{figure*}

Figure \ref{fig:cross_sections} shows deuteron-induced cross sections extracted from Nagai \textit{et al.} \cite{nagai} and TENDL-2023 \cite{tendl}. Since the simulated TENDL data underestimate the production cross section of $^{177}$Lu (due to its larger mass number) \cite{chris,akm,lopez}, the $^{176}$Yb($d,p$)$^{177}$Lu excitation function is replaced by an interpolation of the experimental data (figure \ref{fig:cs}) \cite{herman,man} in what we refer to as TENDL-2023$^\dag$. With this modification, the range-averaged cross sections calculated using the precise version of equation \ref{ran_avg} take on similar magnitudes. It is further observed that for $^{177}$Lu,  $\left\langle \sigma \right\rangle$ peaks around 17.9 MeV, which corresponds to the designed accelerator energy.
\vfill

\begin{figure}[htbp]
    \centering
    \hspace*{-0.35cm}
    \includegraphics[width=9.2cm]{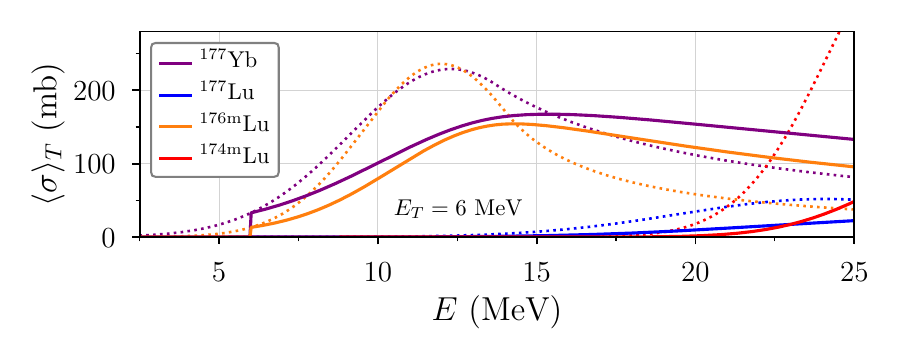}

    \vspace*{-0.75cm}
    \hspace*{-0.35cm}
    \includegraphics[width=9.2cm]{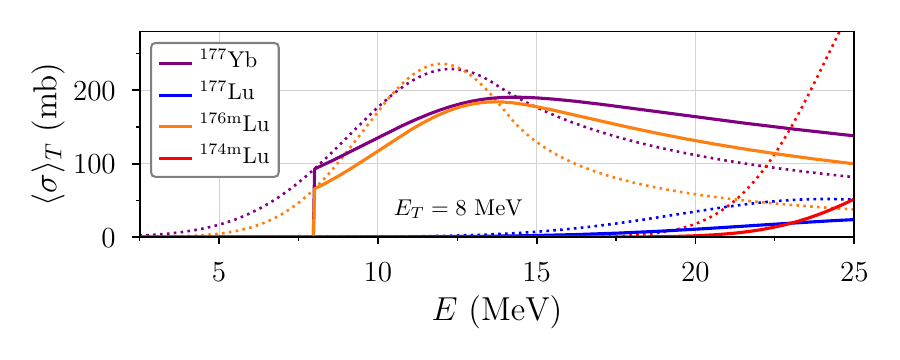}

    \vspace*{-0.75cm}
    \hspace*{-0.35cm}
    \includegraphics[width=9.2cm]{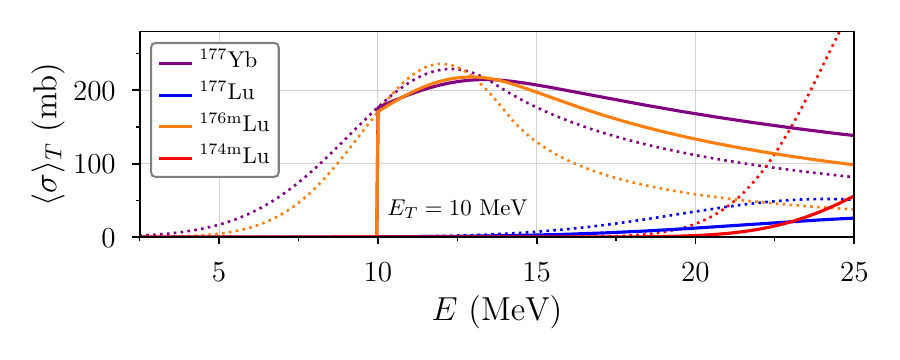}
    
\vspace*{-0.4cm}
\caption{\label{fig:rang_avg2}
Range-averaged cross sections for ions in the intervals $E\rightarrow 6,\,8,\,10$ MeV (i.e. $T<R(E)$), where $E$ is the initial energy of the beam). The calculations are made using excitation functions given by Nagai \textit{et al.} \cite{nagai}.
\vspace*{-0.2cm}
}
\end{figure}

As the stopping power increases for lower energies, the  $^{176}$Yb($d,p$)$^{177}$Yb cross section approaches zero. Therefore, shortening the target's length to less than the range of the incoming deuterons reduces target heating while maintaining similar production efficiency. Since the deuterons now transmit through the target rather than stopping within it, the range-averaged cross section (\ref{ran_avg}) becomes
\begin{equation}
    \left\langle \sigma \right\rangle_{T} = \frac{1}{T} \int_{E_T}^E \frac{\sigma(E')} {S(E')} \,dE',
\label{ran_avg2}
\end{equation}
where $T=R(E)-R(E_T)$ is the target thickness and $E_T$ is the average energy of a deuteron after passing directly through the target \cite{morris}. Figure \ref{fig:rang_avg2} shows $\left\langle \sigma \right\rangle_{T}$ values for target widths corresponding to the intervals $E \rightarrow E_T=6,$ 8, and 10 MeV. When $E=E_T$, we see that $\left\langle \sigma \right\rangle_{T} = \sigma$, as expected. For higher values of $E_T$ (which correspond to a shorter target for a fixed beam energy), the maximum value of $\left\langle \sigma \right\rangle_{T}$ shifts between 17.9 and 12.5 MeV, where $\sigma$ is largest.

Since, for fixed beam energies, the yield of $^{177}$Lu decreases with thinner targets, the beam energy and target thickness should be chosen to optimize production and cost. Table \ref{tab:energy_ranges} considers three beams with fixed energy: 16, 18, and 20 MeV, which all deposit 10 MeV into a Yb$_2$O$_3$ target. Here, $\Delta E /T$ is proportional to the heat absorbed by the target,
$T\langle \sigma \rangle_{T}$ is proportional to the $^{177}$Lu production rate, and $T\langle \sigma \rangle_{T}/E$ is proportional to the production rate per unit of applied beam energy. Although the 20 MeV beam has the highest production rate, the 18 MeV beam has greater production efficiency and saves the cost of an additional DTL.  Compared to the 16 MeV beam, the 18 MeV beam has a higher yield and shifts the deuteron Bragg peak further outside the target, resulting in less energy deposited per unit distance.

\vspace*{0.4cm}
\begin{table}[htpb]
\begin{ruledtabular}
\scalebox{0.95}{
\begin{tabular}{lccc}
 \multicolumn{1}{l}{$E\rightarrow E_T$ (MeV)}&
  $16\rightarrow 6$ & $18\rightarrow 8$ & $20\rightarrow 10$ \\
  \colrule
  \rule{0pt}{2.8ex}
  \!\!$\Delta E$ (MeV) & $10$ & $10$ & $10$ \\[3pt]
  $T$ (mm) & $0.32$ & $0.36$ & $0.40$ \\[3pt]
  $\Delta E /T$ (MeV/mm) & $31.3$ & $27.8$ & $25.0$ \\[3pt]
  $\langle \sigma \rangle_{T}$ (mb) & $167.4$ & $175.4$ & $167.9$ \\[3pt]
  $T\langle \sigma \rangle_{T}$ (mm-mb) & $53.6$ & $63.1$ & $67.2$ \\[3pt]
  $T\langle \sigma \rangle_{T}/E$ (mm-mb/MeV) & $3.35$ & $3.51$ & $3.36$
\end{tabular}
}
\end{ruledtabular}
\caption{\label{tab:energy_ranges}%
$D^+$ energy and Yb$_2$O$_3$ target constants used to optimize the accelerator design for $^{177}$Yb/$^{177}$Lu production.} 
\end{table}

\vspace*{1mm}

Therefore, the 17.9 MeV $D^+$ design is well suited for optimal $^{177}$Lu  production. For a 0.36 mm Yb$_2$O$_3$
target, most ions are expected to deposit around 10 MeV of their initial kinetic energy. As a result, the target will be cooler than if the full 17.9 MeV were deposited, while the production losses will remain negligible. 
\vfill\eject

To further enhance target cooling, a water-cooled copper-backed plate—approximately 1 mm thick—can be incorporated into the assembly. A related study by Praena \textit{et al.} examined water circulation speeds ranging from 5 to 15 m/s, which are also expected to be suitable for our device \cite{javier}.
\vfill

\section{Production Rates}

The secondary ion flux for a reaction of type $i$, or the number of ions produced per unit radial distance squared per second, $\psi = \psi^{\{i\}}$,  is given by
\begin{eqnarray}
     \psi = 
     \int \Phi(E^*)  \!\int\,\frac{\Sigma(E')}{S(E')} \, dE'\, dE^*
\label{Nr_1}
\end{eqnarray}
where $\Phi$ is the deuteron beam flux, $\Sigma = N\sigma$ is the macroscopic cross section for the reaction, and $N$ is the number density of Yb atoms in the target. Using the range-averaged cross section defined in equations \ref{ran_avg} and \ref{ran_avg2}, we can express the second integral in equation \ref{Nr_1} as
\begin{eqnarray}
    \int \,\frac{\Sigma(E')}{S(E')} \, dE' =  R(E^*)\left\langle \Sigma  \right\rangle\!(E^*) \,&&H(E_T-E^*) \nonumber \\
    + \,T\left\langle \Sigma  \right\rangle_{T}\!(E^*) \,&&H(E^*-E_T)
\label{int_sim}
\end{eqnarray}
where the Heaviside step functions $H(E)$ denote the boundary conditions for ions that stop in the target and those that pass through it. Since our design uses a thin target through which the beam fully traverses, the first part of equation \ref{int_sim} vanishes, resulting in a formula that does not explicitly depend on $R$.

For a Gaussian beam profile, the beam flux can be expressed as the product of the integrated flux $\phi$ and a Gaussian shape function $g$:
\begin{equation}
    \Phi(E^*) = \phi \, g(E^*)
\end{equation}
where $\phi= \epsilon\frac{J}{qa}$, $\epsilon$ is the transmission efficiency of the beam, $J$ is the current, $q$ is the charge, and $a$ is the area. In our case, we assume that $a$ is equal to the frontal target area and that $\eta=0.985$ when $J=10$ mA.  Since the standard deviation of the beam energy is small relative to the overall beam energy (see figure \ref{phase}), and because $\sigma$ changes little under these variations, we have $g(E^*) \sim \delta(E^*-E)$, as for a purely mono-energetic beam. Hence, equation \ref{Nr_1} becomes
\begin{eqnarray}
    \psi &\simeq&  \phi\, T  \!\int \delta(E^*-E) \left\langle \Sigma \right\rangle_{T}\!(E^*) \,dE^* \nonumber \\
    &=&  \phi\, T \,N \left\langle \sigma \right\rangle_{T}\!(E).
\label{Nr_2}
\end{eqnarray}
The proportionality of the constants listed in table \ref{tab:energy_ranges} is now evident.
Due to imperfect collimation, angle variations will cause some ions to have a longer target track, which means that equation \ref{Nr_2} will slightly underestimate the actual production rate. To account for this and for generality, we subsequently consider calculations over the interval $18 \rightarrow8$ MeV, which will differ only slightly from the designed 17.9 MeV accelerator.

For a circular Yb$_2$O$_3$ target with a radius of $r=1$ cm, enriched to 99\% $^{176}$Yb ($\eta=0.99$), table \ref{tab:N_r} lists isotope production rates, calculated using $\Gamma=\pi r^2 \psi$, or
\begin{eqnarray}
    \Gamma = \eta\,\epsilon\, \frac{J}{q} \, T \,N \left\langle \sigma \right\rangle_{T}\!(E).
\label{Nr_3}
\end{eqnarray}
 Compared to a pure Yb target ($T=0.541$ mm), the Yb$_2$O$_3$ target ($T=0.359$ mm) produces fewer $^{177}$Yb atoms—$1.065 \times 10^{13}$ compared to $1.367 \times 10^{13}$ atoms/s—when using the Nagai data. However, the range-averaged production rate of the Yb$_2$O$_3$ target exceeds that of pure Yb—$9.428 \times 10^{13}$ compared to $8.038 \times 10^{13}$ atoms/cm$^3$/s. The rates derived from the Nagai and TENDL$^\dag$ data are similar overall, particularly when using the interpolated experimental cross sections \cite{herman,man} to calculate $^{177}$Yb production. The direct production rate of $^{177}$Lu differs by about a factor of 1.6 between the data sets. However, this difference has little consequence, since the $^{177}$Yb production rate is 18–28 times greater.

\begin{table}[htpb]
\begin{ruledtabular}
\scalebox{0.95}{
\begin{tabular}{lcc}
 \multicolumn{1}{l}{$\Gamma^{\{i\}}$ ($\times 10^{13}$atom/s)}&
  Nagai & TENDL-2023$^\dag$ \\
  \colrule
  \rule{0pt}{2.8ex}
  \!\!\!\! $^{176}$Yb$(d,p)^{177}$Yb & $1.065$ & $1.070$  \\[2pt]
  $^{176}$Yb$(d,n\gamma)^{177m}$Lu & n.a. & $2.122 \times 10^{-2}$  \\[2pt]
  $^{176}$Yb$(d,n)^{177g}$Lu & $3.853\times 10^{-2}$ & $6.051 \times 10^{-2}$\\[2pt]
  $^{176}$Yb$(d,2n\gamma)^{176m}$Lu & $0.902$ & $1.179$\\[2pt]
  $^{176}$Yb$(d,2n)^{176g}$Lu & n.a. & $1.259$\\[2pt]
  $^{176}$Yb$(d,3n)^{175}$Lu & n.a. & $2.885$\\[2pt]
  $^{176}$Yb$(d,4n\gamma)^{174m}$Lu & $1.418\times 10^{-3}$ & $9.837\times 10^{-6}$\\[2pt]
  $^{176}$Yb$(d,4n)^{174g}$Lu & $4.642\times 10^{-2}$ & $3.025\times 10^{-5}$\\[2pt]
  $^{176}$Yb$(d,\text{anything})$ & n.a. & $7.105$
\end{tabular}
}
\end{ruledtabular}
\caption{\label{tab:N_r}%
Ion formation rate for 10 mA, 18 MeV $D^+$ on a 0.36 mm Yb$_2$O$_3$ target ($18 \rightarrow 8$ MeV) calculated using data from Nagai \cite{nagai} and TENDL$^\dag$ \cite{tendl}, the latter of which includes $^{176}$Yb($d,p$)$^{177}$Yb data interpolated from \cite{man,herman}. Activations leading to stable Yb isotopes are excluded from the table.} 
\end{table}

\begin{figure}[htbp]
    \centering
    \hspace*{-0.28cm}
    \includegraphics[width=9.1cm]{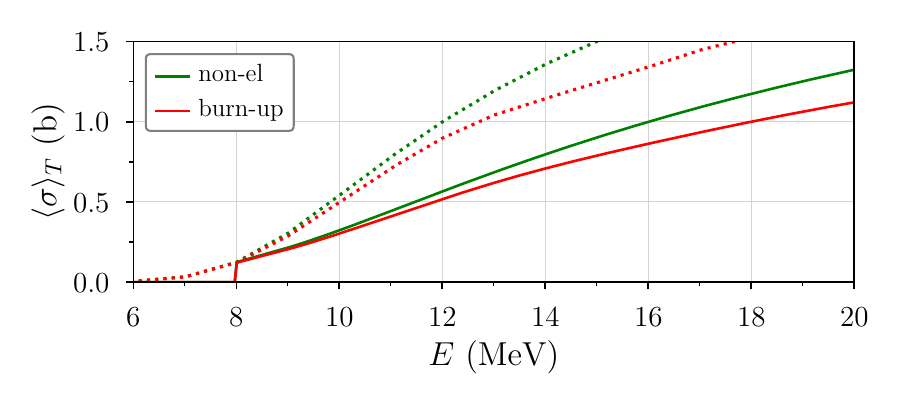}
    
    \vspace*{-0.35cm}
    \hspace*{-0.29cm}
    \includegraphics[width=9.2cm]{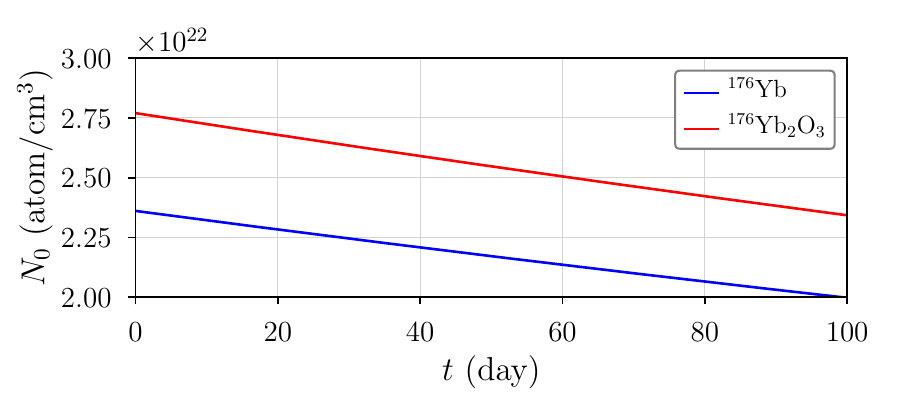}
\vspace*{-0.7cm}
\caption{\label{BU_tar} 
\textit{top.} Non-elastic and target burn-up cross sections for $^{176}$Yb. \textit{bottom.} Disintegration of $^{176}$Yb in Yb and Yb$_2$O$_3$ targets over a 100-day period.}
\end{figure}

For stable isotopes not considered by Nagai \textit{et al.}, the simulated TENDL data allow us to estimate values that can be used for isotopic abundance and target burn-up calculations. The non-elastic cross section for deuterons can be expressed as
\begin{eqnarray}
    \sigma^\text{non}(d) &=& \sum_{0 \leq \nu_n,\,\nu_p,\dots} \!\!\!
 \sigma (  d, \!\!\!\!\!\sum_{q=n,\,p,\dots }  \!\!\!\!\! \nu_{q} \, q  ) \\[5pt]
 &=& \sigma(d,n) + \sigma(d,2n) + \dots + \sigma(d,p) + \sigma(d,2p) \nonumber \\
 && \;\;\;\; +\dots +\sigma(d,n+p) + \sigma(d,2n+p) + \dots \nonumber
\end{eqnarray}
where $\nu_q$ indicates the number of particles of type $q$ produced in a single reaction. Since $^{176}$Yb($d,d'$) and $^{176}$Yb($d,n+p$) reactions retain $^{176}$Yb atoms in the target, the target burn-up cross section is given by
\begin{eqnarray}
    \sigma^*(d) &=& \sigma^\text{non}(d) - \sigma(d,d') - \sigma(d,n+p).
\end{eqnarray}
Figure \ref{BU_tar} shows the non-elastic and burn-up cross sections for deuterons on Yb for $E_T=8$ MeV, with ordinary cross sections represented by dotted lines and range-averaged cross sections by solid lines.  For $E=18$ MeV, $\left\langle\sigma\right\rangle_T^{\text{non}}$ and $\left\langle\sigma\right\rangle_T^*$ differ by less than 20\%, resulting in minimal effect on calculations. For a pure Yb target, the total reaction rate is $9.118 \times 10^{13}$ reactions/s, compared to $ 7.123 \times 10^{13}$ reactions/s for Yb$_2$O$_3$—seven times that of $^{177}$Yb production. The burn-up rates are $7.771 \times 10^{13}$ reactions/s for Yb and $ 6.070 \times 10^{13}$ reactions/s for Yb$_2$O$_3$, respectively. Overall, it is projected to take more than 60 days of continuous irradiation to disintegrate more than 10\% of $^{176}$Yb in the target. This suggests that recycled targets can be reused multiple times \cite{yang}.

\section{Yield Calculations}

Since the produced radionuclides have short half-lives, isotope decay during and after irradiation must be considered as a function of time. The average density of a reaction rate in the target is given by dividing the reaction flux by the target thickness, $\psi/T = \phi \,N \langle \sigma \rangle$. This quantity represents the number of reactions of type $i$ per cubic centimeter. For indirect production of $^{177}$Lu through $^{176}$Yb($d,p$)$^{177}$Yb reactions, we have
\begin{eqnarray*}
    \frac{dN_{^{176}\text{Yb}}}{dt} &=& -\phi \langle \sigma \rangle^*_{^{176}\text{Yb}} N_{^{176}\text{Yb}} \\ 
    \frac{dN_{^{177}\text{Yb}}}{dt} &=& \phi \langle \sigma \rangle_{^{177}\text{Yb}} N_{^{176}\text{Yb}} \\&-& \left(\lambda_{^{177}\text{Yb}} + \phi \langle \sigma \rangle_{^{177}\text{Yb}}^*  \right)N_{^{177}\text{Yb}} 
    \\ 
    \frac{dN_{^{177}\text{Lu}}}{dt} &=&  \lambda_{^{177}\text{Yb}} N_{^{177}\text{Yb}} \\ &-& \left(\lambda_{^{177}\text{Lu}} + \phi \langle \sigma \rangle_{^{177}\text{Lu}}^*  \right)N_{^{177}\text{Lu}}.
\end{eqnarray*}
Here, $\langle \sigma \rangle_{^{176}\text{Yb}}^*=\langle \sigma \rangle_{^{176}\text{Yb},\,T}^*$ represents the average target burn-up cross section, while $\langle \sigma \rangle_{^{177}\text{Yb}}$ and $\langle \sigma \rangle_{^{177}\text{Yb}}^*$ correspond to the production and burn-up of $^{177}$Yb. The decay constant $\lambda_{^{177}\text{Yb}}$ leads to indirect production of $^{177}$Lu, which further decays into stable $^{177}$Hf.
Generalizing the first-order differential equations for gain and loss terms, $G$ and $L$, we can write them as
\begin{eqnarray}
    \dot{N_0} &=& -L_0 N_0 \nonumber\\ 
    \dot{N_1} &=& G_1 N_0 - L_1 N_1\nonumber\\ 
    \dot{N_2} &=& G_2 N_1 - L_2 N_2 \nonumber.
\end{eqnarray}
The equations are then solved by a modified version of Bateman's formula \cite{jerzy},
\begin{equation}
    N_n(t) = N_{0,T} \left(\prod_{j=1}^{n} G_j \right)\sum_{k=0}^{n} \,e^{-L_k t} \!\!\!\!\!\prod_{l=0;\,l\neq k}^n \!\!\! (L_l - L_k)^{-1}
\label{BM_1}
\end{equation}
where $N_{0,T} = N_{T}(t=0)$ is the initial number density of the target and $N_n=N_n^{\{i\}}$ is the $n$-th decaying radionuclide produced by the reaction. As shown in figure \ref{BU_tar}, burn-up of the $^{176}$Yb target ($n=0$) is given by 
\begin{equation}
    N_{^{176}\text{Yb}}(t) = N_{0,^{176}\text{Yb}} \,e^{- \phi \langle \sigma \rangle_{^{176}\text{Yb}}^* t}
\end{equation}
where $t$ is the irradiation time. In contrast to neutron irradiations, where target burn-up is significant \cite{knapp}, burn-up can be neglected in accelerator-based production calculations without making much difference. For indirect 
$^{177}$Lu production Without target burn-up, we have
\begin{eqnarray*} 
    \frac{dN_{^{177}\text{Yb}}}{dt} &=& \phi \langle \sigma \rangle_{^{177}\text{Yb}} N_{^{176}\text{Yb}} - \lambda_{^{177}\text{Yb}} N_{^{177}\text{Yb}} 
    \\ 
    \frac{dN_{^{177}\text{Lu}}}{dt} &=& \lambda_{^{177}\text{Yb}} N_{^{177}\text{Yb}} - \lambda_{^{177}\text{Lu}} N_{^{177}\text{Lu}}, 
\end{eqnarray*}
or,
\begin{eqnarray}
    \dot{N_1} &=& \phi \langle \sigma \rangle_1 N_0-\lambda_1 N_1 \nonumber\\ 
    \dot{N_2} &=& \lambda_1 N_1 - \lambda_2 N_2 \nonumber ,
\end{eqnarray}
where the gain and loss terms now correspond only to the initial production rate and decay constants.
Equation \ref{BM_1} now becomes
\begin{equation}
    N_n(t) =\frac{\phi \langle \sigma \rangle_1 N_{0}}{\lambda_n} \,\sum_{j=1}^{n} \,\left(1-e^{-\lambda_j t}\right) \!\!\!\!\prod_{k=1;\,k\neq j}^n \!\!\! {\lambda_k}\,(\lambda_k - \lambda_j)^{-1}.
\label{BM_2}
\end{equation}
Hence, for $^{176}$Yb($d,p$)$^{177}$Yb reactions, we have ($n=1$)
\begin{equation}
    N_{^{177}\text{Yb}}(t) = \frac{\phi \langle \sigma \rangle_{^{177}\text{Yb}} N_{^{176}\text{Yb}}}{\lambda_{^{177}\text{Yb}}} \left(1- e^{- \lambda_{^{177}\text{Yb}} t} \right)
\end{equation}
and ($n=2$)
\begin{eqnarray}
    N_{^{177}\text{Lu}}(t) &=& \frac{\phi \langle \sigma \rangle_{^{177}\text{Yb}} N_{^{176}\text{Yb}}}{\lambda_{^{177}\text{Lu}}} \left(1- e^{- \lambda_{^{177}\text{Lu}} t} \right) \nonumber\\
    &&+\,\frac{\phi \langle \sigma \rangle_{^{177}\text{Yb}} N_{^{176}\text{Yb}}}{\lambda_{^{177}\text{Lu}}-\lambda_{^{177}\text{Yb}}} \left(e^{- \lambda_{^{177}\text{Lu}} t} - e^{- \lambda_{^{177}\text{Yb}} t} \right). \nonumber\\
\end{eqnarray}
Similarly, for $^{176}$Yb($d,n$)$^{177}$Lu reactions, we have ($n=1$)
\begin{equation}
    N_{^{177}\text{Lu}}(t) = \frac{\phi \langle \sigma \rangle_{^{177}\text{Lu}} N_{^{176}\text{Yb}}}{\lambda_{^{177}\text{Lu}}} \left(1- e^{- \lambda_{^{177}\text{Lu}} t} \right).
\end{equation}
The overall yield is then found by summing over all of the contributing reactions: $N_{^{177}\text{Lu}}(t) = N_{^{177}\text{Lu}}^{\{d,p\}}(t)+N_{^{177}\text{Lu}}^{\{d,n\}}(t)$,
or generally, for an isotope $I$,
\begin{equation}
    N_I(t) = \sum_{i} \delta_{nI}\, N_n^{\{i\}}.
\label{N_sum}
\end{equation}

\begin{figure}[htbp]
    \centering
    \hspace*{-0.35cm}
    \includegraphics[width=9.3cm]{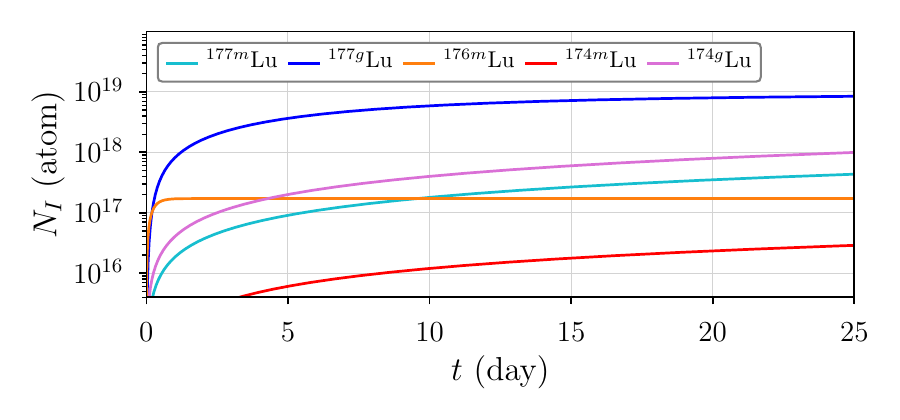}
    
    \vspace*{-0.35cm}
    \hspace*{-0.37cm}
    \includegraphics[width=9.3cm]{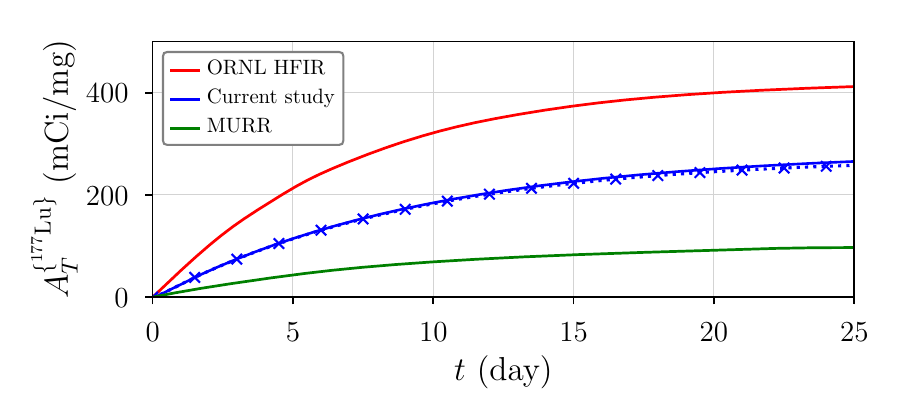}
\vspace*{-0.6cm}
\caption{\label{fig:comparison} 
\textit{top.} Radioisotope yields for long irradiations of 10 mA, 18 MeV $D^+$ ions on a 99\% enriched [$^{176}$Yb]Yb$_2$O$_3$ target. \textit{bottom.} Comparison of the resulting Yb target specific activity with reactor-based production data extracted from reference \cite{dash}. The x-markers indicate when target burn-up is included.}
\end{figure}

Figure \ref{fig:comparison} shows the number of lutetium atoms in the target for long irradiations, with all possible reactions summed according to equation \ref{N_sum}. Due to its short half-life ($t_{1/2} = 3.66$ d), the production of $^{176\text{m}}$Lu rapidly reaches saturation, while $^{177}$Lu increases gradually for decaying $^{177}$Yb. The bottom plot shows the specific activity of $^{177}$Lu (mCi/mg of $^{176}$Yb target) projected by our design, compared to that from neutron irradiation as reported in other studies \cite{dash}.

Although the main advantage of the accelerator-based production method is its higher radiopurity, its overall yield is also competitive with reactor-based methods. Here, the lower production cross section for the deuterons is offset by their higher flux—$1.99\times 10^{16}$ $d$/cm$^2$/s on the target compared to $2\times 10^{15}$ $n$/cm$^2$/s for the Oak Ridge High Flux Isotope Reactor (ORNL HFIR) and $1\times 10^{14}$ $n$/cm$^2$/s for the Missouri University Research Reactor (MURR). For our result, x-markers represent target burn-up given by equation \ref{BM_1}, which results in only a marginal difference.

\begin{figure}[htbp]
    \centering
    \hspace*{-0.33cm}
    \includegraphics[width=9.2cm]{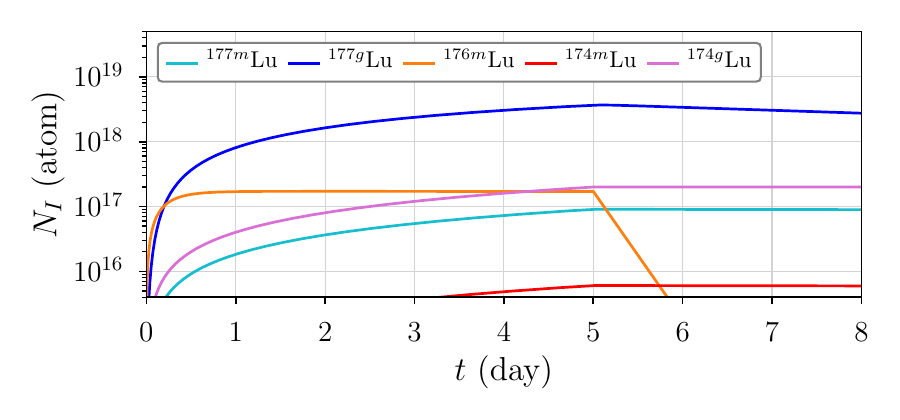}
    
    \vspace*{-0.35cm}
    \hspace*{-0.22cm}
    \includegraphics[width=9.2cm]{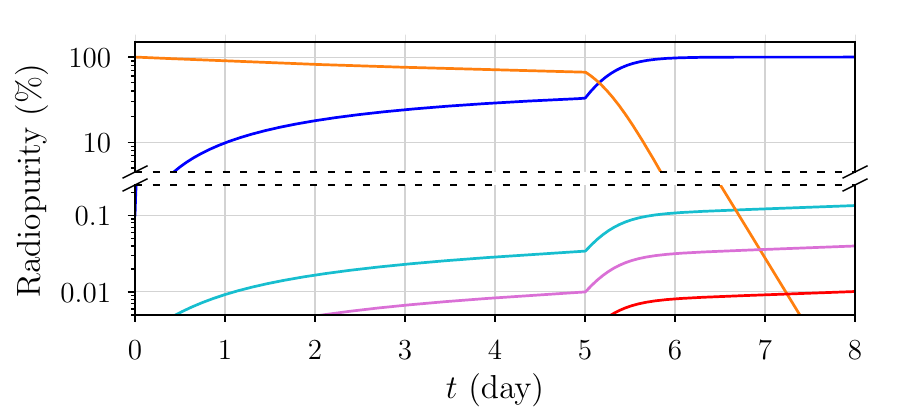}
\vspace*{-0.6cm}
\caption{\label{5day} 
\textit{top.} Lutetium isotope yields and \textit{bottom.} corresponding radiopurity after a 5-day irradiation.}
\end{figure}

Since the activity of $^{177}$Lu begins to saturate after a few days, shorter irradiation times are more effective. For an irradiation that ends at time $t=t'$, we can express the remaining number of an $m$-th decay product as
\begin{equation}
    N_{I,m} (\tau)=\frac{1}{\lambda_m} \sum_{j=1}^m \,N'_{I,j} \sum_{k=j}^{m} \,\lambda_k\, e^{-\lambda_k \tau} \!\!\!\!\!\prod_{l=j;\,l\neq k}^m \!\!\! {\lambda_l}\,(\lambda_l - \lambda_k)^{-1}
\end{equation}
where $\tau = t-t'$ is the time elapsed after irradiation and $N'_{I,j}$ is the concentration of the $j$-th product at $t'$. For $^{177}$Lu ($m=2$), we have
\begin{eqnarray}
    N_{^{177}\text{Lu}}(\tau) &=&  N'_{^{177}\text{Lu}} \,e^{- \lambda_{^{177}\text{Lu}} \tau} \nonumber \\[5pt]
    &&+\,  \frac{N'_{^{177}\text{Yb}}\, \lambda_{^{177}\text{Yb}}}{\lambda_{^{177}\text{Lu}}-\lambda_{^{177}\text{Yb}}} \left(e^{- \lambda_{^{177}\text{Yb}} \tau} - e^{- \lambda_{^{177}\text{Lu}} \tau} \right).
    \nonumber\\
\end{eqnarray}
Figure \ref{5day} now shows lutetium radioisotope yields after a 5-day irradiation. As before, the number of $^{176m}$Lu atoms quickly saturate the target but now decay away rapidly after the irradiation is stopped. 

To determine the radiopurity of an isotope $I$, we divide its activity $\alpha_I(t)=\lambda_I N_I(t)$ by the sum of all activities of the same nuclide. Although the radiopurity of $^{176m}$Lu initially exceeds that of $^{177}$Lu—67.14\% compared to 32.82\%—it is quickly surpassed. After a 2-day processing time, the radiopurity of $^{177}$Lu is 99.81\%, while that of $^{176m}$Lu is just 0.028\%.  For a 12-day irradiation and 2-day processing time, the radiopurity of each isotope is nearly the same.  For longer processing times, the radiopurity of $^{174m}$Lu and $^{177m}$Lu increase since their half-lives are longer than that of $^{177}$Lu. However, the activity of these longer-lived isotopes is generally negligible within the time frame during which the product can be used. After one half-life of $^{177}$Lu, or 6.647 days following a 5-day irradiation, the radiopurity of $^{177}$Lu remains above 99.73\%.

\begin{table}[htpb]
\begin{ruledtabular}
\scalebox{0.95}{
\begin{tabular}{lcccc}
 \multicolumn{1}{l}{$t=5$ d:}&
  $N_i$ ($\times 10^{18}$atoms) & $M_i$ (mg) & $A_i$ (GBq) & $P_i$ (\%) \\
  \colrule
  \rule{0pt}{2.8ex}
  \!\!\!\! $^{177m}$Lu & $0.0907$ & $0.027$ & $4.54$ & $0.034$  \\[2pt]
  
  $^{177g}$Lu & $3.65$ & $1.073$ & $4406$ & $32.82$  \\[2pt]
  
  $^{176m}$Lu & $0.172$ & $0.05$ & $9015$ & $67.14$  \\[2pt]
  
  $^{174m}$Lu & $0.0061$ & $0.002$ & $0.342$ & $0.003$  \\[2pt]
  
  $^{174g}$Lu & $0.02$ & $0.058$ & $1.328$ & $0.01$  \\[4pt]
  $\tau=48$ hr: \\
  
  \colrule 
  \rule{0pt}{2.8ex}
  \!\!$^{177m}$Lu & $0.0899$ & $0.026$ & $4.50$ & $0.12$  \\[2pt]
  
  $^{177g}$Lu & $3.05$ & $0.897$ & $3682$ & $99.81$  \\[2pt]
  
  $^{176m}$Lu & $2.0\times 10^{-5}$ & $6\times 10^{-6}$ & $1.027$ & $0.028$  \\[2pt]
  
  $^{174m}$Lu & $0.006$ & $0.002$ & $0.338$ & $0.009$  \\[2pt]
  
  $^{174g}$Lu & $0.02$ & $0.058$ & $1.326$ & $0.036$  \\[2pt]

\end{tabular}
}
\end{ruledtabular}
\caption{\label{tab:N_n_5day}%
Lutetium isotope yields after a 5-day irradiation and 48-hour processing time.} 
\vspace{0.2cm}
\begin{ruledtabular}
\scalebox{0.95}{
\begin{tabular}{lcccc}
 \multicolumn{1}{l}{$t=12$ d:}&
  $N_i$ ($\times 10^{18}$atoms) & $M_i$ (mg) & $A_i$ (GBq) & $P_i$ (\%) \\
  \colrule
  \rule{0pt}{2.8ex}
  \!\!\!\! $^{177m}$Lu & $0.214$ & $0.063$ & $10.72$ & $0.064$  \\[2pt]
  
  $^{177g}$Lu & $6.50$ & $1.91$ & $7840$ & $46.47$  \\[2pt]
  
  $^{176m}$Lu & $0.172$ & $0.05$ & $9015$ & $53.44$  \\[2pt]
  
  $^{174m}$Lu & $0.0142$ & $0.004$ & $0.807$ & $0.005$  \\[2pt]
  
  $^{174g}$Lu & $0.48$ & $0.14$ & $3.181$ & $0.019$  \\[4pt]
  $\tau=48$ hr: \\
  
  \colrule 
  \rule{0pt}{2.8ex}
  \!\!$^{177m}$Lu & $0.213$ & $0.062$ & $10.63$ & $0.16$  \\[2pt]
  
  $^{177g}$Lu & $5.36$ & $1.58$ & $6469$ & $99.76$  \\[2pt]
  
  $^{176m}$Lu & $2.0\times 10^{-5}$ & $6\times 10^{-6}$ & $1.027$ & $0.016$  \\[2pt]
  
  $^{174m}$Lu & $0.014$ & $0.004$ & $0.8$ & $0.012$  \\[2pt]
  
  $^{174g}$Lu & $0.48$ & $0.138$ & $3.177$ & $0.049$  \\[2pt]

\end{tabular}
}
\end{ruledtabular}
\caption{\label{tab:N_n_12day}%
Lutetium isotope yields after a 12-day irradiation and 48-hour processing time.} 
\end{table}

Tables \ref{tab:N_n_5day} and \ref{tab:N_n_12day} list the yield and activity of each radioisotope. From our 1.04 g Yb$_2$O$_3$ target design, 1.07 mg of $^{177}$Lu can be obtained after 5 days of irradiation. Producing 1.9 mg of $^{177}$Lu—77\% more than the initial amount—requires an additional 7 days of irradiation, a 140\% increase in irradiation time. Optimal irradiation times are thus left to be determined by target processing costs. Due to saturation, the concentrations of $^{177}$Yb and $^{176m}$Lu are not affected by longer irradiations, while the production of $^{174m}$Lu and $^{177m}$Lu is insignificant.

For direct $^{177}$Lu production by irradiating an 82\% [$^{176}$Lu]LuCl$_3$ with thermal neutrons ($\phi_n = 1.2 \times 10^{14}$ n/cm$^2$/s), Sairanbayev \textit{et al.} estimate activities ranging from 364 to 537 GBq/mg for irradiations lasting between 6.25 and 17 days \cite{sairan}. The corresponding ratio of $^{177m}$Lu to $^{177}$Lu activities was found to be less than 0.025. In our study, this ratio is $1.03 \times 10^{-3}$ for 5-day and $1.37 \times 10^{-3}$ for 12-day irradiations. After a 5-day irradiation and 2-day processing time, the specific activity of $^{177}$Lu, averaged over the mass of all radionuclides, is 3.75 TBq/mg. Considering that long-lived $^{176}$Lu and stable $^{175}$Lu are also produced by ($d,2n$) and ($d,3n$) reactions, the specific activity averaged over the total lutetium mass is found to be 594 GBq/mg.

After irradiation, the Yb$_2$O$_3$ target can be dissolved in a 4 M solution of hydrochloric acid \cite{RA}. Macro amounts of Yb are separated from the solution to isolate the $^{177}$Lu product, which is ultimately precipitated into LuCl$_3$. 84–95\% of Lu can be retained in this process \cite{yang} with Yb impurities less than 10$^{-4}\%$ \cite{RA}. As the overall processing only takes several hours, our design is expected to meet the requirement that radiochemical purity exceed 99\%.  

\vfill

\section{Conclusions}

Currently, $^{177}$Lu treatments cost around \$14,500 per dose and require at least 4 doses in total \cite{vogel,morgan}. Given that these doses are administered every 6 to 8 weeks, it is necessary to localize production to ensure a reliable quantity. Compared to reactor-based production methods, accelerators offer several advantages, such as easier supervision, enhanced safety, and lower maintenance costs \cite{yiwei}.  In addition, accelerators produce less nuclear waste and are capable of yielding a greater radiopurity of the desired isotope. 

In this study, we demonstrated the feasibility of accelerator-based $^{177}$Lu production and introduced new methods, based on experimental and simulated excitation functions, that can be used to optimize future accelerator designs. This includes the calculation of the target-averaged cross section $\langle \sigma \rangle_T$ and its corresponding relations, as provided in table \ref{tab:energy_ranges}.

In conclusion, we presented the design of a 10 mA, 17.9 MeV $D^+$ linac intended to irradiate a 99\% enriched [$^{176}$Yb]Yb$_2$O$_3$ target. We showed that our design is capable of producing 4.4 TBq of $^{177}$Lu after 5 days and 6.4 TBq after 12 days, which meets the dose requirements for thousands of patients. The minimal burn-up of the target suggests that remaining Yb$_2$O$_3$ may be recycled and reused in multiple irradiations. Compared to traditional neutron-based production methods, our design achieves a competitive $^{177}$Lu yield (1 mg in the first 5 days) and exceptionally high radiopurity ($>$99.8\%). This design is expected to play an integral role in the future production of $^{177}$Lu and other radioisotopes for the treatment of metastatic lesions and neuroendocrine cancers.